# Electrical Impedance Response of Gamma Irradiated Gelatin based Solid Polymer Electrolytes Analyzed Using a Generalized Calculus Formalism

Tania Basu[a], Abhra Giri[a], Sujata Tarafdar[a], Shantanu Das[b]

[a]*Department of Physics*, Jadavpur University, Kolkata-700032, West Bengal, India.
[b]*RCSDS Reactor Control Division*, Bhabha Atomic Research Centre, Mumbai-400085, Maharastra, India

*Corresponding author email: tania16basu@gmail.com .
Phone: +91 33 24146666 (extn. 2760), Fax: +91 33 24148917

*Abstract*— The electrical impedance response of Gelatin based solid polymer electrolyte to gamma irradiation is investigated by impedance spectroscopy. An analysis based on Poisson-Nernst-Plank model, incorporating fractional time derivatives is carried out. A detailed derivation for anomalous impedance function is given.The model involves boundary conditions with convolution of the fractional time derivative of ion density and adsorption desorption relaxation kinetics. A fractional diffusion-drift equation is used to solve the bulk behavior of the mobile charges in the electrolyte. The complex adsorption-desorption process at the electrode-electrolyte interface produces an anomalous effect in the system. The model gives a very good fit for the observed impedance data for this biopolymer based solid electrolyte in wide range of frequencies. We have compared different parameters based upon this model for both irradiated and unirradiated samples.

*Keywords*—*Caputo fractional derivative; solid electrolyte; gamma irradiation; memory kernel; convolution; adsorption-desorption.*

## 1. INTRODUCTION

Impedance spectroscopy provides a detailed picture of how charge carriers behave in a conducting system, when analyzed through a realistic model. We report results on a solid polymer electrolyte with gelatin as the main constituent. The process of conduction may be divided into three basic phenomena; the first one is diffusion-drift phenomena, which can be a Fickian or non Fickian process governed by an integer order time derivative or fractional order time derivative in the diffusion equation. The non-integer order derivatives (fractional derivatives) in the rate of diffusing species implies that the process is non-Markovian, and with a memory [1]. The non-Fickian nature of diffusion starting from Catteneo's diffusion (1948) is described in detail in [1]. The impedance spectroscopy data as obtained in these experiments may be related to anomalous diffusion in the bulk material due to spatial disorder in the solid electrolyte matrix. The spatial disorder leads to non-Debye relaxation which manifests itselfas temporal fractional derivatives [1, 2, 3, 4]. Fig. 1, describes pictorially, the heterogeneity in the bulk electrolyte.

The structure is heterogeneous on micro-scales but appears homogeneous on scales much larger than the heterogeneities. This is one of the plausible reasons of ions in the bulk electrolyte (in this case solid polymer) obeying fractional diffusion equation at the small time scales and normal integer order diffusion equation at larger time scale. The process of transport may be mixed as well, i.e. governed by anomalous and normal diffusion taking place simultaneously.

The second process is that of local charge separation giving rise to a potential obtained from Poisson's equation. We use it to get the potential profile in the bulk electrolyte and thus the electric field function. We then apply Gauss's law at the boundary electrodes to get the surface charge density and thereby the total charges at the surface; from there we derive the rate of change of the total charge, giving the current function. The potential at the electrode and the current at the electrodes give the impedance function.

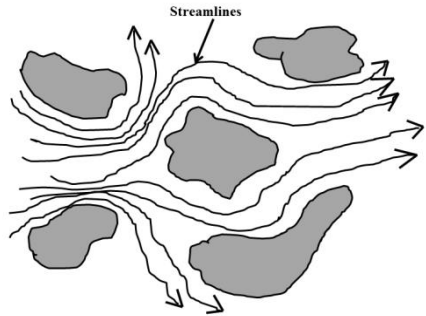

Fig.1 The disorder in diffusing path which manifests as fractional time derivative in the diffusion expression

The third phenomena which is very important is at the surfaces (i.e. the two electrodes), this is the kinetics of adsorption-desorption which gives the current density at the boundary - related to convolution of 'memory kernel' - the adsorption-desorption kinetics with (fractional) rate of change of ion density. With the memory kernel as zero we get the blocking electrode case - i.e. the classical case where the current at the boundary is zero.

The anomalous response is at low frequencies, where it is observed that the real part of the impedance increases with decrease in frequency. This can be explained by adding fractional order impedance in the calculations [4, 5, 6, 7]. By taking the electrode (brass) electrolyte impedance as,

$$Z_i(\omega) = C_q (i\omega)^{-q}$$

Where $C_q > 0; \quad 0 < q < 1$

We can explain the growth of impedance as frequency decreases. This interfacial impedance contributes to the total resistance of the cell the term,

$$\mathrm{Re}\{Z_i\} = \mathrm{Re}\{C_q(i\omega)^{-q}\} = C_q \omega^{-q} \cos(q\pi/2)$$

which is frequency dependent and particularly as $\omega \to 0$ it diverges, that is one way to explain the anomalous impedance observed. But in the interface impedance the manifestation of fractional order $q$ comes through the roughness of the electrode surface (as observed in super-capacitors) [4]; whereas in our case the brass electrodes are smooth, without noticeable roughness. The other way to explain this anomalous behavior is to have different mobility for anions and cations and therefore different diffusion constants [8],

$$\mathbb{D}_+ = 10 \mathbb{D}_-$$

But we are dealing with ions having same mobility and the diffusion constants. The ions in this case are principally $H_3O^+$ and $OH^-$. With normal integer order diffusion have a strong adsorption condition at the electrodes [9], to explain the anomalous impedance we introduce the fractional order diffusion in the bulk as well as at the interface of the electrodes.

## 2. MATERIALS AND METHOD OF MEASUREMENT

Gelatin films with formaldehyde as cross linking antifungal agent and different wt% of glycerol as plasticizer are prepared by solution cast method [10]. The masses of glycerol used for the different samples were 0.25, 0.50, 0.75, 1.00, 1.25, 1.35 and 1.50 x$10^{-3}$ kg (corresponding to weight percent of plasticizer 10.00, 18.18, 25.00, 30.77, 35.71, 37.50 and 40.00% respectively). The transparent films (thickness ~ 500 μm) are then exposed to gamma chamber ($^{60}Co_{\gamma rays}$) at the Department of Food Technology, Jadavpur University and UGC -

DAE CSR, Kolkata centre with doses (20kGy, 40kGy, 60kGy, 80kGy, 100kGy) at the rate of 6.4 kGy/hr and 3.4 kGy/hr respectively. An Agilent LCR meter (E4980A precision meter) was used to measure the complex impedance Z at room temperature (30°C) in the frequency range from 20 Hz to 2 MHz.

## 3. Experimental Results

The ion-conductivity σ (dc) obtained from the Cole–Cole plots are shown in Fig. 2 as function of the plasticizer fraction without irradiation (A) and with irradiation (B) and (C). The dc conductivity σ (dc) increased by four orders of magnitude with the addition of 10 wt % glycerol, σ (dc) is maximum for 35.71 wt% glycerol ~ 9.14 x$10^{-3}$ S/m at room temperature (30°C) without irradiation (Fig. 2 (A)) [2]. After irradiation the conductivity of the sample containing 35.71 wt% glycerol is seen to decrease by one order of magnitude and we obtained the maximum conductivity at 60 kGy dose ~ 9.63 x$10^{-4}$ S/m at room temperature (30°C) (Fig.2 (C)). For higher doses the dc conductivity falls off as same as earlier work [10, 11]. In the absence of added salt, the charge carriers were assumed to be primarily Hydronium ions ($H_3O^+$) [12]. There may have been a small percentage of impurity ions, such as P, S, Ca, Cl and N coming from the gelatin (Merck, 99.9% pure). Energy-dispersive X-ray spectroscopy results before irradiation [10] and after irradiation which are not displayed in this work indicated the presence of slight amounts of C, O, N, P, S, Ca and Cl in the polymer matrix. We will assume that the major phenomena is due to univalent Hydronium (positive ion) and Hydroxide (OH negative ion); and their mobility is same, and so is their diffusion coefficient $\mathbb{D}$. We neglect the effect of trace impurity ions.

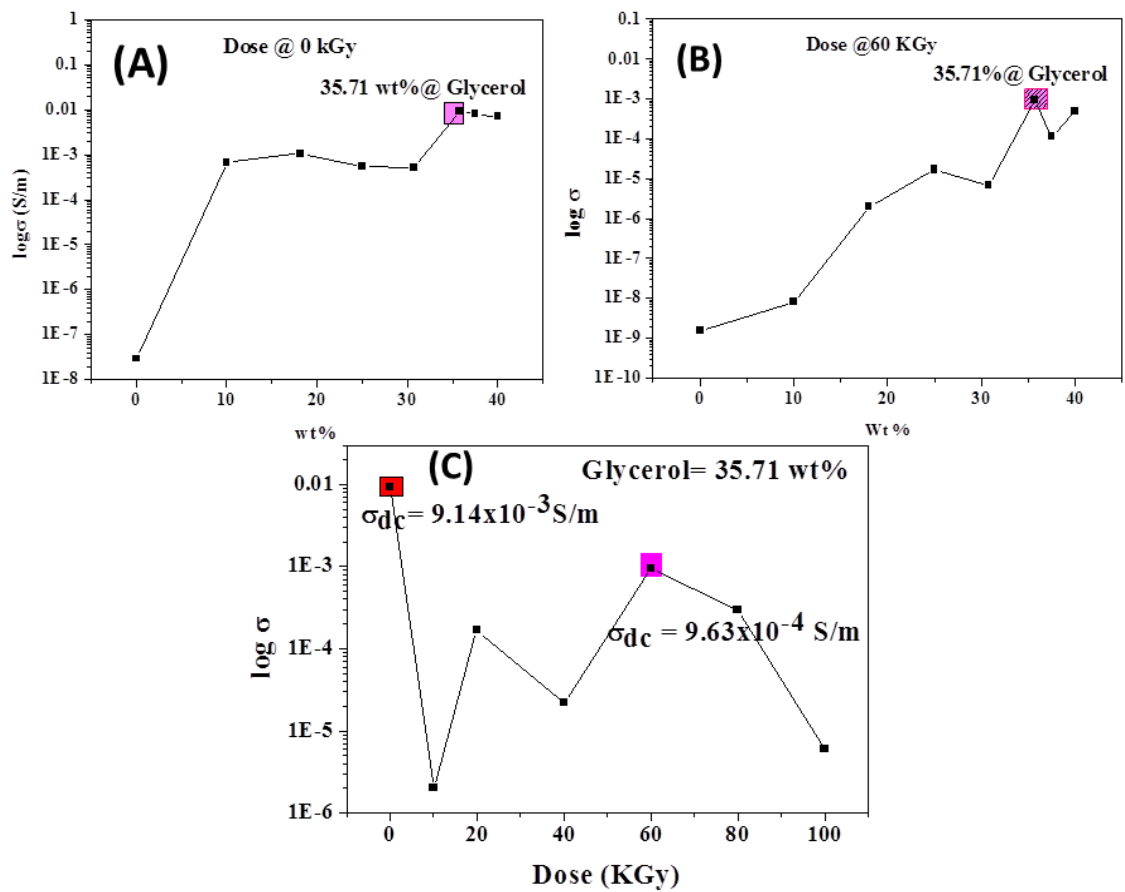

Fig.2 Log σ vs. different wt% of glycerol (A) at 0kGy (B) 60kGy. (C) Log σ vs. different doses for 35.71 wt% of glycerol.

# 4. THEORY

## 4.1. Poisson-Nernst-Plank (PNP) Model

In our present work we are trying to compare the behavior of gelatin based samples before and after gamma irradiation with the help of anomalous diffusion analysis which is based on Poisson-Nernst-Plank model (modified by introducing fractional time derivatives for diffusing species). This formalism was developed by Lenzi et al [13-22]. This theory is developed for a liquid electrolyte such as Nematic Crystals, ultra pure water; but it is applicable to solid electrolyte as well. The ion conduction mechanism in non-crystalline solid electrolyte is very similar to liquids, in particular for polymer samples above glass transition where dynamic disorder is present. Such materials, though apparently solid, are characterized by segmental motion, with parts of the macromolecular chain in incessant motion on very small spatial and temporal scales [23], leaving the center of mass of the molecule stationary.

The model developed by Lenzi et al. considers ion transport to be entirely through anomalous diffusion. We modify their theory assuming that some fraction of the charge carriers ($h$) diffuse through normal admittance and rest ($1-h$) through anomalous admittance. The material consists of a uniform "solvent" with, equally charged positive and negative ions moving with same mobility. The solvent is assumed to have an effective dielectric constant $\varepsilon$ for particular plasticizer content. The sample with surface area $S$ and thickness $d$ is placed between the brass electrodes. The anomalous response is at low frequencies, where it is observed that the real part of the impedance increases with decrease in frequency.

This phenomena does not get reflected by the usual (integer order) Poisson-Nernst-Plank (PNP) [15, 17, 20, 24] theory with boundary condition of blocking electrodes. The PNP theory visualizes constant and flat real impedance at low frequency, with a high frequency cut-off like a true low pass filter. This approach fails to reproduce the experimentally observed rise of real part of impedance function with decrease in frequency.

The PNP is integer order differential equation theory. Perhaps this is inadequate to model the complex mechanism. Memorized relaxation (a non-Debye process) and complex adsorption-desorption processes at the electrodes (complex surface effects) are reckoned to play an important role in the observed impedance spectra.

# 5. MATHEMATICAL DEVELOPMENT OF THE MODEL

## 5.1. Constitutive equation for bulk electrolyte with generalized calculus

We write the constitutive for bulk electrolyte as follows,

$$\int_0^1 dq \left( k(q) \right) \frac{\partial^q}{\partial t^q} n_\alpha(z,t) = -\frac{\partial}{\partial z} j_\alpha(z,t) \qquad q \in (0,1) \tag{1}$$

This (1) is fractional diffusion equation with distributed order where $n_\alpha(z,t)$ represents the bulk number density of diffusing and drifting species $\alpha = +$ for positive species and $\alpha = -$ for negative species, and $j_\alpha(z,t)$ is current flux. Equation (1) is a generalization of the normal integer order constitutive equation of diffusion, which is the following,

$$\frac{\partial}{\partial t} n_\alpha(z,t) = -\frac{\partial}{\partial z} j_\alpha(z,t)$$

which we get considering the kernel $k(q) = \delta(q-1)$ in (1). The kernel $k(q)$ is a distribution function which is in general a continuous function of the fractional order $q$ between zero and one. In our diffusion case we assume a spread of fractional order between zero and one; this is introduced in our integration in (1), [1, 25, 26].

The fractional order imbibes memory [1, 27, 28] in relaxation of $n_\alpha(z,t)$. The second constitutive equation is current density is given by,

$$j_\alpha(z,t) = -\mathbb{D} \frac{\partial}{\partial z} n_\alpha(z,t) \mp \frac{q_e \mathbb{D}}{k_B T} n_\alpha(z,t) \frac{dV}{dz} \tag{2}$$

$\mathbb{D}$ is diffusion coefficient, $V$ is the excitation voltage, $T$ is the ambient temperature, $k_B$ is Boltzmann constant, $q_e$ is the electronic charge.

The fractional derivative of the order $\bar{q}$ is Caputo's derivative. This definition of fractional derivative is requiring that $n_\alpha(z,t)$ be a differentiable function in the interval of interest at all points. The definition of Caputo derivative [1] is,

$$^C_{t_0}D^{q_0}_t n_\alpha(z,t) = \frac{\partial^{q_0}}{\partial t^{q_0}} n_\alpha(z,t) \triangleq \frac{1}{\Gamma(k-\bar{q})} \int_{t_0}^{t} dt'(t-t')^{-q_0+k-1} n_\alpha^{(k)}(z,t') \quad (3)$$

$$k-1 < q_0 < k, \quad k \in Z^+, \quad q_0 \in \Re^+$$

The

$$n_\alpha^{(k)} \equiv \frac{\partial^k}{\partial t^k} n_\alpha(z,t)$$

is the $k$-th integer order derivative (just greater than real order $q$). We use Caputo's fractional derivative instead of classical Riemann-Liouvelli (RL)[13]. The Caputo derivative requires differentiability condition where as the RL derivative requires function to be continuous need not be differentiable. RL derivative of a constant is not zero, but a decaying power function. If the process of differentiation is starting at start point of minus infinity the RL derivative of constant is zero. We use the property of Caputo derivative in the derivation with perturbation i.e. $^C_{t_0}D^{q_0}_t(N) = 0$, where $N$ is constant number; as demonstrated in detail in subsequent sections for impedance calculations.

Taking $k(q) = a\delta(q-1) + b\delta(q-q_0)$ in (1) and with (2) we obtain the fractional diffusion-drift equation as follows:

$$a\frac{\partial}{\partial t}n_\alpha(z,t) + b\frac{\partial^{q_0}}{\partial t^{q_0}}n_\alpha(z,t) = -\frac{\partial}{\partial z}j_\alpha(z,t) = \mathbb{D}\frac{\partial^2 n_\alpha(z,t)}{\partial z^2} \pm \frac{q_e\mathbb{D}}{k_B T}n_\alpha(z,t)\frac{d^2 V}{dz^2}$$

The above is a diffusion-drift equation with order one with weight $a$ and fractional order $q_0 \in (0,1)$, with weight $b$.

The dimension of $a$ is nil i.e. dimensionless, and $b$ has the dimension of $[time]^{q_0}$ [20]. If we have small perturbation of harmonic nature for voltage excitation say $V(z,t) = \phi(z)e^{i\omega t}$ then the ion density too will have perturbation as $n_\alpha(z,t) = N + \delta n_\alpha(z,t)$; where $N$ is constant, and $\delta n_\alpha(z,t) = n_\alpha(z)e^{i\omega t}$ for ions we can write the above diffusion-drift equation as follows:

$$[a(i\omega) + b(i\omega)]n_\alpha(z) = \mathbb{D}\left[n_\alpha''(z) \pm \frac{q_e N}{k_B T}\phi''(z)\right]$$

The double prime at RHS is double derivative w.r.t. $z$. With $a=1$ and $b=0$ we have normal integer order diffusion equation, and putting perturbed values we obtain:

$$i\omega n_\alpha(z) = \mathbb{D}_e\left[n_\alpha''(z) \pm \frac{q_e N}{k_B T}\phi''(z)\right]$$

Clearly if, in the integer order system, we replace, $\mathbb{D}_e$ by $\mathbb{D}_e \equiv \mathbb{D}/\left[a + b(i\omega)^{q_0-1}\right]$ we obtain the generalized diffusion-drift equation with kernel $k(q) = a\delta(q-1) + b\delta(q-q_0)$. The diffusion constant has a property of frequency dependency. For integer order diffusion equation with $a=1$ and $b=0$, we have frequency independent case and $\mathbb{D}_e = \mathbb{D}$. But for a pure fractional order case with fractional order $q_0$ we get a frequency dependent diffusion constant. Putting $a=0$ and $b=1$, we get $\mathbb{D}_e = \mathbb{D}/(i\omega)^{q_0-1}$ which is for a pure fractional order diffusion-drift equation. This $\mathbb{D}_e$ takes its form as the case may be with the diffusion-drift equation in the overall impedance function.

### 5.2. Memory Integral & Fractional Derivative and generalization of Diffusion-Drift equation with memory integral

We write the time evolution of a dynamic system as,

$$\frac{\partial}{\partial t}n_\alpha(t) = -\int_0^t d\bar{t}\, K_D(t-\bar{t})n_\alpha(\bar{t}) = -K_D(t) * n_\alpha(t)$$

Above represents memory integral [1] i.e. all instances for $\bar{t}=0$ to $\bar{t}=t$ contribute to situation at present time. This above is convolution of memory kernel $K_D(t)$ and the relaxing quantity i.e. $n_\alpha(t)$. A Markovian case is relaxation without memory. Say we have kernel $K_D(t) = \delta(t)/\tau_D$.

We have the integer order diffusion-drift equation with its frequency domain expression as follows, with $n_\alpha(z,t) = n_\alpha(z)e^{i\omega t}$ and $V(z,t) = \phi(z)e^{i\omega t}$.

$$\frac{\partial}{\partial t} n_\alpha(z,t) = \mathbb{D}_e \frac{\partial}{\partial z}\left[\frac{\partial n_\alpha(z,t)}{\partial z} \pm \frac{q_e N}{k_B T}\frac{dV}{dz}\right]$$

$$(i\omega)n_\alpha(z) = \mathbb{D}_e\left[n_\alpha''(z) \pm \frac{q_e N}{k_B T}\phi''(z)\right]$$

If we add memory kernel for a relaxation of $\frac{\partial}{\partial t} n_\alpha(t) = -\int_{-\infty}^{t} d\bar{t}\, K_D(t-\bar{t}) n_\alpha(\bar{t})$ then we have consolidated expression with memory integral as memory based diffusion-drift equation as:

$$\frac{\partial}{\partial t} n_\alpha(z,t) = \mathbb{D}\frac{\partial}{\partial z}\left[\frac{\partial n_\alpha(z,t)}{\partial z} \pm \frac{q_e N}{k_B T}\frac{dV}{dz}\right] - \int_{-\infty}^{t} d\bar{t}\, K_D(t-\bar{t}) n_\alpha(z,\bar{t})$$

Note that we have purposely used $\mathbb{D}$ instead of $\mathbb{D}_e$. The above has frequency domain expression as noted below with $n_\alpha(z,t) = n_\alpha(z)e^{i\omega t}$ and $V(z,t) = \phi(z)e^{i\omega t}$.

$$(i\omega)n_\alpha(z) = \mathbb{D}\left[n_\alpha''(z) \pm \frac{q_e N}{k_B T}\phi''(z)\right] - \{\bar{K}_D(i\omega)\}n_\alpha(z)$$

$$(i\omega)n_\alpha(z)\left[1 + \frac{\{\bar{K}_D(i\omega)\}}{(i\omega)}\right] = \mathbb{D}\left[n_\alpha''(z) \pm \frac{q_e N}{k_B T}\phi''(z)\right]$$

Where $\{\bar{K}_D(i\omega)\}$ is frequency Fourier transformed of the time domain memory kernel $K_D(t)$. With this memory kernel we are getting a frequency dependent diffusion constant as,

$$\mathbb{D}_e = \frac{\mathbb{D}}{[1 + \{\bar{K}_D(i\omega)\}/(i\omega)]}$$

Here we note that the diffusion constant needs to be frequency dependent when we invoke the memory kernel, in the presence of fractional time derivatives in the diffusion-drift equation. We had earlier obtained $\mathbb{D}_e \equiv \mathbb{D}/\left[a + b(i\omega)^{q_0-1}\right] = \mathbb{D}/\left[1 + \{(i\omega)^{q_0}\}/(i\omega)\right]$ in case of $a = b = 1$. A similar result is obtained with memory convolution;

$$\frac{\partial}{\partial t} n_\alpha(z,t) + \frac{\partial^{q_0}}{\partial t^{q_0}} n_\alpha(z,t) = \mathbb{D}\frac{\partial^2 n_\alpha(z,t)}{\partial z^2} \pm \frac{q_e \mathbb{D}}{k_B T} n_\alpha(z,t) \frac{d^2 V}{dz^2}$$

Thus for cases with fractional order time derivative we have relaxation with memory which is a non-Markovian process in the diffusion-drift equation. Indeed the usual diffusion-drift equation is an approximation only valid in time scales that are large when compared with the time scales in which diffusion-causing collision takes place. With the fractional derivative part absent in the diffusion-drift equation and only integer order time derivatives present, we have a situation of infinite phase velocity of information propagation, with zero collision time, which is rather non-physical. The presence of the fractional derivatives with $\bar{q} < 1$ introduces some kind of damping, making finite phase velocity, implying finite collision time in a physical scenario [1].

### 5.3. The simple adsorption-desorption at the boundary interface and Langmuir approximation

The adsorption-desorption phenomenon is the surface dynamics where the electrolyte's ions get attached and de-attached while in contact with the surface electrode. We will consider physic-sorption, the species adsorbing retains its identity unlike chemisorption where a chemical bond is formed [29]. Adsorption requires a particle to lose energy during collision with the surface. Some of the particles bombarding the surface will bounce back off the surface. But at any particular density certain fraction will remain giving rise to some coverage of the surface. The covering ratio or coverage is defined as [29],

$$\sigma_R = \frac{\text{Number of surface sites occupied}}{\text{Total number of surface sites}} = \frac{\sigma}{\sigma_0}$$

The rate of adsorption is proportional to ρ and the number of adsorbing sites at the surface i.e.

$$\frac{d\sigma}{dt} = k_a \rho(\sigma_0 - \sigma)$$

Where, $k_a$ [m/s] is the rate constant for adsorption; $\rho$ [m$^{-1}$] is the bulk density (in one dimension) of adsorbate just in front of adsorbing surface, and $\sigma_0 - \sigma$ is the total number of the free sites at the surface. $\sigma$ and $\sigma_0$ are dimensionless. The rate of desorption is proportional to number of adsorbed species, that is

$$\frac{d\sigma}{dt} = -k_d \sigma$$

where the term $k_d \left[ (s)^{-1} \right]$ is the rate constant for desorption. Introducing reduced quantities as $\sigma_R = \sigma/\sigma_0$ and $\rho_R = \rho/\rho_0$ [29]. At the equilibrium "net rate of adsorption" i.e. adsorption rate plus desorption rate is zero that is,

Net rate of adsorption $= k_a \rho(\sigma_0 - \sigma) + (-k_d \sigma) = 0$

From above we get Langmuir isotherm as $\sigma_R = (\alpha \rho_R)/(1 + \alpha \rho_R)$. The term $\alpha$ governs the steady state whence $\alpha = \kappa \tau \rho_0 / \sigma_0$ with $\tau = 1/k_d$ and $\kappa = k_a \sigma_0$. The characteristic time constant $\tau$ for desorption and $\kappa$ are associated with adsorption phenomena. Notice that $\kappa \tau$ has dimension of length. Also we have from the balance equation $\rho_R = (\sigma_R)/[\alpha(1-\sigma_R)]$. Here we assume that the boundary electrodes in the $x-y$ plane are situated at $z = -d/2$ and at $z = +d/2$, for a one dimensional case. We indicate the density of the particles of the medium in the position $z$ at a given time $t$, and $\sigma(t)$ is the surface density by $\rho(z,t)$. The equilibrium values of these are

$$\lim_{t \to \infty} \rho(z,t) = \rho(z) = \rho \qquad \lim_{t \to \infty} \sigma(t) = \sigma$$

The net rate of adsorption (adsorption rate + desorption rate) is as follows:

$$\frac{d\sigma}{dt} = k_a \rho(\sigma_0 - \sigma) - k_d \sigma$$

$$\frac{d\sigma}{dt} = \kappa \rho \left(1 - \frac{\sigma}{\sigma_0}\right) - \frac{1}{\tau}\sigma$$

$$\frac{d\sigma_R}{dt} = \kappa \frac{\rho_0}{\sigma_0} \rho_R (1 - \sigma_R) - \frac{1}{\tau}\sigma_R$$

At the equilibrium $d\sigma_R/dt = 0$, one gets the Langmuir isotherm, which has thickness dependence. In fact, since $\rho_0 d$ is the initial number of particles per unit area, the conservation of number of particles at any time requires following,

$$2\sigma(t) + \int_{-d/2}^{d/2} \rho(z,t)dz = \rho_0 d$$

At the equilibrium $t \to \infty$ thus we have $2\sigma + \rho d = \rho_0 d$. From this equilibrium relation we obtain $\sigma_R + \alpha \delta(\rho_R - 1) = 0$; with $\delta = d/2\kappa\tau$ as dimensionless thickness, where $\alpha = \kappa \tau \rho_0 / \sigma_0$.

We have adsorption dynamics as derived just above as,

$$\frac{d\sigma}{dt} = \kappa \rho \left(1 - \frac{\sigma}{\sigma_0}\right) - \frac{1}{\tau}\sigma$$

In the limit $\sigma \ll \sigma_0$ adsorbed sites are very small compared to total number of sites; we get the adsorption dynamics as following (along with its dimensionless representation)

$$\frac{d\sigma}{dt} + \frac{1}{\tau}\sigma = \kappa \rho \qquad \frac{d\sigma_R}{dt} + \frac{1}{\tau}\sigma_R = \frac{\kappa \rho_0}{\sigma_0} \rho_R$$

The Green's function if we choose as $\sigma_g(t) = \kappa e^{-t/\tau}$ is solution to the above (homogeneous) differential equation, and this we call the Langmuir approximation which we will subsequently use. The Langmuir method has given a dynamic expression relating surface adsorbed species to the species density of bulk. The particular solution to this will be convolution of the Green's function with $\rho(z,t)$ appearing at RHS of the Langmuir dynamic equation. So we can write the solution as,

$$\sigma(z,t) = \int_{-\infty}^{t} d\bar{t} [\kappa e^{-(t-\bar{t})/\tau}][\rho(z,\bar{t})]$$

The above is rather similar to memory integral where memory kernel $K_m(t) = \kappa e^{-t/\tau}$. The LHS of the above equation is proportional to the surface current density $\sigma(z,t)|_{z=\pm d/2} \sim j_\alpha(z,t)|_{z=\pm d/2}$. Now the density of diffusing ions inside the bulk material $\rho(z,t)|_{z=\pm d/2} \sim n_\alpha(z,t)|_{z=\pm d/2}$. With this we may write boundary condition (of a particular type as) as:

$$j_\alpha(z,t)\big|_{z=\pm\frac{d}{2}} = \pm\int_{-\infty}^{t} d\bar{t}\,\kappa\, e^{-(t-\bar{t})/\tau} n_\alpha(z,\bar{t})\,\big|_{z=\pm d/2}$$

The frequency domain translation of the above is,

$$j_\alpha(z,s)\big|_{z=\pm\frac{d}{2}} = \pm\kappa\left(\frac{\tau}{1+s\tau}\right)(n_\alpha(z,s))\big|_{z=\pm d/2}$$

For steady state frequency response put complex Laplace frequency $s = \text{Re}\{s\} + i\omega = i\omega;\quad \text{Re}\{s\} = 0$ to get,

$$j_\alpha(z,i\omega)\big|_{z=\pm\frac{d}{2}} = \pm\kappa\left(\frac{\tau}{1+i\omega\tau}\right)(n_\alpha(z,i\omega))\big|_{z=\pm d/2}$$

$$j_\alpha(z)\big|_{z=\pm\frac{d}{2}} = \pm\left(\frac{\kappa\tau}{1+i\omega\tau}\right)(n_\alpha(z))\big|_{z=\pm d/2}$$

The next section further generalizes the boundary condition by involving fractional time derivative in the rate for number density.

*5.4. Unusual boundary condition at the electrodes*

For the influence of the surface, on the ions, we have to consider the boundary conditions subjected to the expression of current density at the electrode boundary $z = \pm d/2$;

$$j_\alpha(z,t)\big|_{z=\pm\frac{d}{2}} =$$

$$= \pm\int_0^1 d\bar{q}\,(\bar{k}(\bar{q})) \times \int_{-\infty}^{t} d\bar{t}\,(K_m(t-\bar{t}))\frac{\partial^{\bar{q}}}{\partial \bar{t}^{\bar{q}}} n_\alpha(z,\bar{t})\,\big|_{z=\pm\frac{d}{2}}$$

$$\bar{q} \in (0,1]$$

(4a)

This is an unusual process of dynamics of adsorption-desorption happening at the boundary, a much generalized way to write the process is with the help of generalized calculus.

For a blocking electrode, the memory kernel of the convolution at RHS of (4a) is, $K_m(t) = 0$ gives boundary condition as $j_\alpha(z,t)\big|_{z=\pm d/2} = 0$. For the fractional order distribution function as $\bar{k}(\bar{q}) = \delta(\bar{q}-1)$ and the memory kernel as at RHS as delta function i.e. $K_m(t) = \delta(t)$ we return to a very simple case of adsorption-desorption at the boundary without any memory (Markovian-case), namely

$$j_\alpha(z,t)\big|_{z=\pm d/2} = \pm\frac{\partial}{\partial t} n_\alpha(z,t)\bigg|_{z=\pm d/2}$$

The surface effect given by above integer order differential equation is true if the process at the electrodes is taking place as if, at all conditions the $n_\alpha$'s (the ions) are detached or attach to the surface uniformly, and with memory-less dynamics - in an homogeneous and smooth background. In reality the adsorption energy of one site may not be independent of a neighboring one's occupational state [29] with adsorption happening at inner layers too. The relaxation times of adsorption and desorption may not have any set average. The distribution may be diverging with power-law statistics [1]. This uncertain wait-time statistics may be another cause of anomalous adsorption-desorption at the boundary. Also all the adsorbing sites at boundary may not be equivalent [29]. These are the anomalous conditions at the boundary, and thus a generalized way to express boundary condition is given in (4a).

The adsorption and desorption processes at the surface with Langmuir approximation is employed for this unusual dynamics. The memory kernel with Langmuir approximation i.e. $K_m(t) = \kappa_1 e^{-t/\tau_1}$ gives [20] short range correlation whereas in the case of Delta function as $K_m(t) = \delta(t)$, there is no memory effect.

Therefore, including the memory kernel proposed by Lenzi et al [20] the boundary condition can be modified more precisely including both $h$ fraction for normal and $(1-h)$ fraction for anomalous situations, as described below:

$$h:\quad j_\alpha(z,t)\big|_{z=\pm\frac{d}{2}} = \int_{-\infty}^{t} d\bar{t}\,\kappa_N\, e^{-(t-\bar{t})/\tau_N} \frac{\partial n_\alpha(z,\bar{t})}{\partial \bar{t}}\,\big|_{z=\pm\frac{d}{2}} \qquad (4b)$$

$$(1-h): \quad j_\alpha(z,t)|_{z=\pm\frac{d}{2}} = \int_{-\infty}^{t} d\bar{t}\, \kappa_A e^{-(t-\bar{t})/\tau_A} \frac{\partial^{\bar{q}} n_\alpha(z,\bar{t})}{\partial \bar{t}^{\bar{q}}}\bigg|_{z=\pm\frac{d}{2}} \quad (4c)$$

The precise nature and the origin of anchoring and 'anchoring energy' at surface due to adsorption in these electrolyte cells of impedance spectroscopy are: "Still subject of many fundamental and experimental studies and cannot be considered as solved problem"[29].

*5.5. The boundary operator $\Phi$ for normal & anomalous cases*

First, from the above boundary condition we can obtain the boundary operator in frequency domain $(i\omega)$, for both normal and anomalous situations using (4b) and (4c)

$$j_\alpha(z,t)|_{z=\pm\frac{d}{2}} = \pm \int_0^1 d\bar{q}\,(\bar{k}(\bar{q})) \times \int_{-\infty}^{t} d\bar{t}\,(K_m(t-\bar{t})) \frac{\partial^{\bar{q}}}{\partial \bar{t}^{\bar{q}}} n_\alpha(z,\bar{t})\bigg|_{z=\pm\frac{d}{2}} \quad n_\alpha(z,t) = n_\alpha(z)e^{i\omega t}$$

$$= \pm \int_0^1 d\bar{q}\,(\bar{k}(\bar{q})) \times \int_{-\infty}^{t} d\bar{t}\,(K_m(t-\bar{t}))(i\omega)^{\bar{q}} e^{i\omega \bar{t}} n_\alpha(z)\bigg|_{z=\pm\frac{d}{2}}$$

$$= \pm \Phi(i\omega) n_\alpha(z) e^{i\omega t}$$

Here we have used Caputo derivative ${}_{-\infty}^{C}D_t^q e^{\beta t} = \beta^q e^{\beta t}$ for $\beta > 0$; $0 < q < 1$, thus we write in above ${}_{-\infty}^{C}D_t^{\bar{q}} n_\alpha(z) e^{i\omega t} = n_\alpha(z)(i\omega)^{\bar{q}} e^{i\omega t}$

Define boundary operator as:

$$\Phi(i\omega) = \pm e^{-i\omega t} \int_0^1 d\bar{q}\,(\bar{k}(\bar{q})) \times \int_{-\infty}^{t} d\bar{t}\,(K_m(t-\bar{t}))(i\omega)^{\bar{q}} e^{i\omega \bar{t}}.$$

Taking Langmuir kernel $K_m = \kappa e^{-t/\tau}$ i.e. $K_m(t-\bar{t}) = \kappa e^{-(t-\bar{t})/\tau}$ and placing it above, we have boundary operator as:

$$\Phi(i\omega) = \pm \kappa\tau \left(\frac{1}{1+i\omega\tau}\right) \int_0^1 d\bar{q}\,(\bar{k}(\bar{q}))(i\omega)^{\bar{q}}$$

For normal case the order distribution function $\bar{k}(\bar{q}) = m\delta(\bar{q}-1)$ with Langmuir kernel as $K_{mN} = \kappa_N e^{-t/\tau_N}$, the boundary operator takes the form:

$$\Phi_N(i\omega) = \pm \kappa_N \tau_N \left(m \frac{i\omega}{1+i\omega\tau_N}\right)$$

For anomalous case the order distribution function $\bar{k}(\bar{q}) = n\delta(\bar{q}-\bar{q}_0); 0 < \bar{q}_0 < 1$ and Langmuir relaxation kernel as $K_{mA} = \kappa_A e^{-t/\tau_A}$, the boundary operator takes the form:

$$\Phi_A(i\omega) = \pm \kappa_A \tau_A \left(n \frac{(i\omega)^{\bar{q}_0}}{1+i\omega\tau_A}\right)$$

Here $\tau_N$ and $\tau_A$ signifies desorption relaxation time for normal and anomalous diffusions. Thus we have here the two different length scales for adsorption-desorption at the boundary, namely $\kappa_N \tau_N$ and $\kappa_A \tau_A$.

*5.6. The Poisson's equation and its solution*

The potential gradient is determined via Poisson's equation;

$$\frac{\partial^2}{\partial z^2} V(z,t) = -\frac{q_e}{\varepsilon}\left[n_+(z,t) - n_-(z,t)\right] \quad (5)$$

Time dependent potential exciting the cell is $V(z,t) = \phi(z)e^{i\omega t}$ and at the boundary electrodes the potential is $V(d/2,t) = \pm(V_0/2)e^{i\omega t}$. Let there be a small perturbation in number density by application of a small oscillating voltage $n_\alpha(z,t) = N + \delta n_\alpha(z,t)$ with linear approximation $\delta n_\alpha(z,t) \ll N$. The perturbed density varies as, $\delta n_\alpha(z,t) = n_\alpha(z)e^{i\omega t}$.

Again let $\psi_+(z) = n_+(z) + n_-(z)$ & $\psi_-(z) = n_+(z) - n_-(z)$

From (1) and (2) we have

$$\int_0^1 dq\,(k(q)) \frac{\partial^q}{\partial t^q} n_\alpha = -\frac{\partial}{\partial z}\left[-\mathbb{D}\frac{\partial n_\alpha}{\partial z} \mp \frac{q_e \mathbb{D}}{k_B T} n_\alpha \frac{dV}{dz}\right]$$

$$= \mathbb{D}\frac{\partial^2 n_\alpha}{\partial z^2} \pm \frac{q_e \mathbb{D}}{k_B T} n_\alpha \frac{d^2 V}{dz^2}$$

Substitute $n_\alpha(z,t) = N + \delta n_\alpha(z,t)$. Recognize $\partial[n_\alpha(z,t)]/\partial t = \partial[\delta n_\alpha(z,t)]/\partial t$ and Caputo's derivative $^C\partial^q N/\partial t^q = 0$. Put $n_\alpha(z,t) = N + \delta n_\alpha(z,t)$ and the perturbed density as $\delta n_\alpha(z,t) = n_\alpha(z)e^{i\omega t}$ to get following

$$\int_0^1 dq(k(q)) \frac{\partial^q[N + \delta n_\alpha(z,t)]}{\partial t^q} = \mathbb{D}\frac{\partial^2[N + \delta n_\alpha(z,t)]}{\partial z^2} \pm \frac{q_e \mathbb{D}}{k_B T}[N + \delta n_\alpha(z,t)]\frac{d^2 V}{dz^2}$$

$$\int_0^1 dq(k(q)) \frac{\partial^q[\delta n_\alpha(z,t)]}{\partial t^q} = \mathbb{D}\frac{\partial^2[\delta n_\alpha(z,t)]}{\partial z^2} \pm \frac{q_e \mathbb{D} N}{k_B T}\frac{d^2 V}{dz^2} \pm \frac{q_e \mathbb{D} \delta n_\alpha(z,t)}{k_B T}\frac{d^2 V}{dz^2}$$

Put in above $\delta n_\alpha(z,t) = n_\alpha(z)e^{i\omega t}$ and $V(z,t) = \phi(z)e^{i\omega t}$ and $\delta n_\alpha(z,t) \ll N$, we get:

$$\int_0^1 dq(k(q)) \frac{\partial^q n_\alpha(z)e^{i\omega t}}{\partial t^q} = \mathbb{D}\frac{\partial^2 n_\alpha(z)e^{i\omega t}}{\partial z^2} \pm \frac{q_e \mathbb{D} N}{k_B T}\frac{d}{dz}\phi(z)e^{i\omega t}$$

In our study we are dealing with steady state response, thus the lower terminal in the integral transformed representation is at minus infinity. Also at the minus infinity the function $e^{\beta t}$ is zero. In this study thus we aren't concerned with the initial conditions at minus infinity and all initial conditions related to differential equations are at rest (zero).

We derived

$$\int_0^1 dq(k(q))(i\omega)^q n_\alpha(z) = \mathbb{D}\frac{\partial^2 n_\alpha(z)}{\partial z^2} \pm \frac{q_e \mathbb{D} N}{k_B T}\frac{d^2}{dz^2}\phi(z) \quad (6)$$

Let us write the operator on LHS of above $\int_0^1 dq(k(q))(i\omega)^q \equiv \Lambda(i\omega)$, and write (6) as

$$\frac{\Lambda(i\omega)}{\mathbb{D}} n_\alpha(z) = \frac{\partial^2}{\partial z^2} n_\alpha(z) \pm \frac{q_e N}{k_B T}\frac{d^2 \phi(z)}{dz^2}$$

Using Poisson's expression:

$$\frac{\partial^2}{\partial z^2} V(z,t) = -\frac{q_e}{\varepsilon}[n_+(z,t) - n_-(z,t)]$$

$$\frac{\partial^2}{\partial z^2} \phi(z)e^{i\omega t} = -\frac{q_e}{\varepsilon}[n_+(z)e^{i\omega t} - n_-(z)e^{i\omega t}]$$

$$\frac{d^2 \phi(z)}{dz^2} = -\frac{q_e}{\varepsilon}[n_+(z) - n_-(z)]$$

Substituting in (6) to get

$$\frac{\Lambda(i\omega)}{\mathbb{D}} n_\alpha(z) = \frac{\partial^2 n_\alpha(z)}{\partial z^2} \mp \frac{q_e^2 N}{\varepsilon k_B T}[n_+(z) - n_-(z)] \quad (7)$$

From (7) we segregate the equation for + and − charges as follows:

$$\frac{\Lambda(i\omega)}{\mathbb{D}} n_+(z) = \frac{\partial^2 n_+(z)}{\partial z^2} - \frac{q_e^2 N}{\varepsilon k_B T}[n_+(z) - n_-(z)]$$

$$\frac{\Lambda(i\omega)}{\mathbb{D}} n_-(z) = \frac{\partial^2 n_-(z)}{\partial z^2} + \frac{q_e^2 N}{\varepsilon k_B T}[n_+(z) - n_-(z)]$$

Rearranging above we get two differential equations

$$\frac{\partial^2 n_+(z)}{\partial z^2} = \frac{\Lambda(i\omega)}{\mathbb{D}} n_+(z) + \frac{q_e^2 N}{\varepsilon k_B T}[n_+(z) - n_-(z)] \quad (8)$$

$$\frac{\partial^2 n_-(z)}{\partial z^2} = \frac{\Lambda(i\omega)}{\mathbb{D}} n_-(z) - \frac{q_e^2 N}{\varepsilon k_B T}[n_+(z) - n_-(z)] \quad (9)$$

Adding (8) and (9) we have

$$\frac{\partial^2[n_+(z) + n_-(z)]}{\partial z^2} = \frac{\Lambda(i\omega)}{\mathbb{D}}[n_+(z) + n_-(z)] \quad (10)$$

$$\frac{d^2}{dz^2}\psi_+(z) = \alpha_+^2 \psi_+(z) \qquad \alpha_+^2 = \frac{\Lambda(i\omega)}{\mathbb{D}}$$

Subtracting (9) from (8) we have

$$\frac{\partial^2[n_+(z) - n_-(z)]}{\partial z^2} = \frac{\Lambda(i\omega)}{\mathbb{D}}[n_+(z) - n_-(z)] + \frac{2q_e^2 N}{\varepsilon k_B T}[n_+(z) - n_-(z)] \quad (11)$$

$$\frac{d^2}{dz^2}\psi_-(z) = \alpha_-^2 \psi_-(z) \qquad \alpha_-^2 = \frac{\Lambda(i\omega)}{\mathbb{D}} + \frac{2q_e^2 N}{\varepsilon k_B T} = \frac{\Lambda(i\omega)}{\mathbb{D}} + \frac{1}{\lambda^2}$$

Consolidating (10) and (11) we obtain 2nd order linear differential equation (not FDE)

$$\frac{d^2}{dz^2}\psi_\pm = \alpha_\pm^2 \psi_\pm \quad (12)$$

$$\alpha_-^2 = \frac{\Lambda(i\omega)}{\mathbb{D}} + \frac{1}{\lambda^2} \tag{13}$$

$$\alpha_+^2 = \frac{\Lambda(i\omega)}{\mathbb{D}} \tag{14}$$

Remember that $\lambda^2 = (\varepsilon k_B T)/(2q_e^2 N)$ gives Debye screening length $\lambda$. This is surface effect of bare sample, in contact with a substrate; where we have selective adsorbed charges on the surface screened by opposite charge giving a charge separation of a distance $\lambda$, from the surface. This is the basic phenomenon of formation of Electric Double Layer Capacity ELDC [29]. This bare screening length gets altered when we have potential in the electrolyte cell. The solution of the ordinary differential equation (12) is $\psi_\pm(z) = c_{\pm 1} e^{\alpha_\pm z} + c_{\pm 2} e^{-\alpha_\pm z}$. We have symmetry in potential distribution about centre as $V(z,t) = -V(-z,t)$. Applying this observation we get:

$$\psi_-(z) = 2c_{-1}\left(\frac{e^{\alpha_- z} - e^{-\alpha_- z}}{2}\right) = 2c_{-1} \sinh(\alpha_- z) \tag{15}$$

From Poisson's expression we have,

$$\frac{d^2 V(z,t)}{dz^2} = -\frac{q_e}{\varepsilon}[n_+(z,t) - n_-(z,t)] \quad V(z,t) = \phi(z)e^{i\omega t}$$

$$n_\alpha(z,t) = n_\alpha(z)e^{i\omega t}$$

$$\frac{d^2}{dz^2}\phi(z) = -\frac{q_e}{\varepsilon}[n_+(z) - n_-(z)] = -\frac{q_e}{\varepsilon}\psi_-(z) = -\frac{q_e}{\varepsilon}[2c_{-1} \sinh(\alpha_- z)] \tag{16}$$

Integrating (16) we get gradient of potential i.e. proportional to electric field in the cell,

$$\frac{d\phi(z)}{dz} = -\frac{q_e}{\varepsilon \alpha_-}[2c_{-1} \cosh(\alpha_- z)] + c_1 \tag{17}$$

Integrating (17) we get potential as a function in electrolyte cell,

$$\phi(z) = -\frac{2q_e}{\varepsilon \alpha_-^2}[c_{-1} \sinh(\alpha_- z)] + c_1 z + c_0$$

At origin the center of cell we have $\phi(z)\big|_{z=0} = \phi(0) = 0$ so from above $c_0 = 0$

$$\phi(z) = -\frac{2q_e}{\varepsilon \alpha_-^2}[c_{-1} \sinh(\alpha_- z)] + c_1 z$$

*5.7. Calculation of electric field charge density and current at the boundary electrodes and the cell impedance function:*

The electric field is the negative gradient of the potential function

$$E(z,t) = -\frac{d}{dz}V(z,t) = -\frac{d}{dt}\phi(z)e^{i\omega t} = -\phi'(z)e^{i\omega t} \quad E(\{d/2\},t) = -\phi'(d/2)e^{i\omega t}$$

We apply Gauss law at the electrode at $+d/2$ connected to potential $+V/2$. Gauss's law states that $\oiint_S E(d/2,t).ds = Q_{total}/\varepsilon$. Where $Q_{total}$ is total charge enclosed in volume as $-\sigma_s.S$. LHS of Gauss law is flux through a closed surface of area $S$ is $E(d/2,t).S$. Therefore $E(d/2,t) = -\frac{\sigma_s}{\varepsilon}$ where $-\sigma_s$ is total surface charge density. So, $\sigma_s = -\varepsilon E(d/2,t) = \varepsilon \phi'(z=d/2)e^{i\omega t}$. Therefore total charge at $z = d/2$ is, $Q_{total} = \sigma_s S = \varepsilon S \phi'(d/2)e^{i\omega t}$. See the Fig. 3 of the pill box.

The current is,

$$I(z,t)\big|_{z=d/2} = \frac{d}{dt}Q_{total} = i\omega \varepsilon S \phi'(d/2)e^{i\omega t}$$

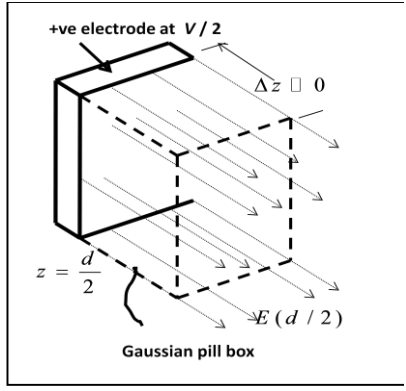

Fig.3 Gaussian pill box at the surface

Impedance is,
$$Z = \frac{V}{I}\bigg|_{z=d/2} = \frac{\phi(d/2)e^{i\omega t}}{i\omega\varepsilon S\phi'(d/2)e^{i\omega t}} = \frac{\phi(d/2)}{i\omega\varepsilon S\phi'(d/2)} \qquad (18)$$

From (2) we have,
$$j_\alpha(z,t) = -\mathbb{D}\frac{\partial}{\partial z}n_\alpha(z,t) \mp \frac{q_e\mathbb{D}}{k_B T}n_\alpha(z,t)\frac{dV}{dz}$$

Using $\Phi(i\omega) = \pm e^{-i\omega t}\int_0^1 d\overline{q}\left(\overline{k}(\overline{q})\right) \times \int_{-\infty}^t d\overline{t}\left(K_m(t-\overline{t})\right)(i\omega)^{\overline{q}}e^{i\omega\overline{t}}$ as obtained earlier we get the following,

$$j_\alpha(z,t) = -\mathbb{D}\frac{\partial}{\partial z}n_\alpha(z,t) \mp \frac{q_e\mathbb{D}}{k_B T}n_\alpha(z,t)\frac{dV}{dz} \qquad n_\alpha(z,t) = N + n_\alpha(z)e^{i\omega t} \qquad V = \phi(z)e^{i\omega t}$$

$$= -\mathbb{D}\frac{d}{dz}n_\alpha(z)e^{i\omega t} \mp \frac{q_e\mathbb{D}N}{k_B T}\frac{d\phi(z)}{dz}e^{i\omega t} \qquad n_\alpha(z) \ll N$$

$$-\mathbb{D}\frac{d}{dz}n_\alpha(z)e^{i\omega t} \mp \frac{q_e\mathbb{D}N}{k_B T}\frac{d\phi(z)}{dz}e^{i\omega t}\bigg|_{z=\pm d/2} = \pm\Phi(i\omega)n_\alpha(z)e^{i\omega t}\bigg|_{z=\pm d/2}$$

$$-\mathbb{D}\frac{d}{dz}n_\alpha(z) \mp \frac{q_e\mathbb{D}N}{k_B T}\frac{d\phi(z)}{dz}\bigg|_{z=\pm d/2} = \pm\Phi(i\omega)n_\alpha(z)\bigg|_{z=\pm d/2}$$

As we segregated the +ve and –ve species earlier for Drift-Diffusion equation (1) and (2) and then adding and subtracting the obtained expression from above we can get;

$$\mathbb{D}\frac{d}{dz}\psi_-(z) + \frac{2q_e\mathbb{D}N}{k_B T}\frac{d\phi(z)}{dz}\bigg|_{z=\pm d/2} = \mp\Phi(i\omega)\psi_-(z)\bigg|_{z=\pm d/2} \qquad (19)$$

$$\mathbb{D}\frac{d}{dz}\psi_+(z)\bigg|_{z=\pm d/2} = \mp\Phi(i\omega)\psi_+(z)\bigg|_{z=\pm d/2} \qquad (20)$$

We have from (15) & (17) $\psi_-(z) = 2c_{-1}\sinh(\alpha_- z)$, and $\frac{d\phi(z)}{dz} = -\frac{q_e}{\varepsilon\alpha_-}\left[2c_{-1}\cosh(\alpha_- z)\right] + c_1$ using them in (19) we get the following,

$$2\mathbb{D}c_{-1}\cosh(\alpha_- z) + \frac{2q_e N\mathbb{D}}{k_B T}\left\{-\frac{2q_e c_{-1}\alpha_{-1}\cosh(\alpha_- z)}{\varepsilon\alpha_-^2} + c_1\right\}\bigg|_{z=\pm\frac{d}{2}}$$

$$= \mp\Phi(i\omega)2c_{-1}\sinh(\alpha_- z)\bigg|_{z=\pm\frac{d}{2}}$$

$$\left[\mathbb{D}\alpha_-\cosh(\alpha_- z) + \frac{q_e N\mathbb{D}}{k_B T}\left(-\frac{2q_e\alpha_-}{\varepsilon\alpha_-^2}\right)\cosh(\alpha_- z) \pm \Phi(i\omega)\sinh(\alpha_- z)\right]c_{-1} + \frac{q_e N\mathbb{D}}{k_B T}c_1\bigg|_{z=\pm\frac{d}{2}} = 0$$

$$\left[\mathbb{D}\alpha_-\cosh(\alpha_- z)\left(1 - \frac{2q_e N}{k_B T\varepsilon\alpha_-^2}\right) \pm \Phi(i\omega)\sinh(\alpha_- z)\right]c_{-1} + \frac{q_e N\mathbb{D}}{k_B T}c_1\bigg|_{z=\pm\frac{d}{2}} = 0$$

The parameters are,
$$\alpha_-^2 = \frac{\Lambda(i\omega)}{\mathbb{D}} + \frac{1}{\lambda^2} \qquad \frac{1}{\lambda^2} = \alpha_-^2 - \frac{\Lambda(i\omega)}{\mathbb{D}}.$$

The impedance dispersion in terms of basic derived operators, use

$$E = \Lambda(i\omega) \pm \alpha_{-}\Phi(i\omega)\tanh\left(\alpha_{-}\frac{d}{2}\right) = \left(\alpha^2 - \frac{1}{\lambda^2}\right)\mathbb{D} \pm \alpha_{-}\Phi(i\omega)\tanh\left(\alpha_{-}\frac{d}{2}\right) \text{ to get}$$

$$Z(i\omega) = \frac{1}{i\omega\varepsilon S \alpha_{-}^2} \left[ \frac{\dfrac{\tanh\left(\alpha_{-}\dfrac{d}{2}\right)}{\lambda^2 \alpha_{-}} + \dfrac{Ed}{2\mathbb{D}}}{1 + \dfrac{\Phi(i\omega)\tanh\left(\alpha_{-}\dfrac{d}{2}\right)}{\alpha_{-}i\omega\lambda^2}\left(\dfrac{i\omega\lambda^2}{\mathbb{D}}\right)} \right] \quad (21)$$

We divide the process by assuming that some species follow normal integer order laws and rest follow the fractional order laws, denoted by fraction $h$ and $(1-h)$ respectively, through admittances $Y_N$ and $Y_A$ to get $Z_{final}$ as,

$$\frac{1}{Z_{final}} = hY_N + (1-h)Y_A = \frac{h}{Z_N} + \frac{1-h}{Z_A} \quad (22)$$

$$Z_N = \frac{1}{i\omega S \varepsilon \alpha_{N-}^2} \frac{\dfrac{\tanh(\alpha_{N-}d/2)}{\lambda^2 \alpha_{N-}} + \dfrac{E_N d}{2\mathbb{D}_N}}{1 + \dfrac{\Phi_N(i\omega)\tanh(\alpha_{N-}d/2)}{\alpha_{N-}i\omega\lambda^2}\left(\dfrac{i\omega\lambda^2}{\mathbb{D}_N}\right)} \quad (23)$$

$$Z_A = \frac{1}{i\omega S \varepsilon \alpha_{A-}^2} \frac{\dfrac{\tanh(\alpha_{A-}d/2)}{\lambda^2 \alpha_{A-}} + \dfrac{E_A d}{2\mathbb{D}_A}}{1 + \dfrac{\Phi_A(i\omega)\tanh(\alpha_{A-}d/2)}{\alpha_{A-}i\omega\lambda^2}\left(\dfrac{i\omega\lambda^2}{\mathbb{D}_A}\right)} \quad (24)$$

$\mathbb{D}_N$ is the normal diffusion coefficient and $\mathbb{D}_A$ is anomalous diffusion coefficient.

## 6. RESULTS AND DISCUSSIONS

We compare the experimental result with the theoretical model (PNP). Introducing the modifications mentioned in section III gives a physically meaningful insight into the behavior of the solid polymer electrolytes (SPEs) under observation.

We calculate the DC conductivity from the Nyquist plot in Fig. 2. The real and imaginary parts obtained from experiments are plotted against the frequency and compared with calculations from the model. The parameters which give the best fit are given in a tabular form (Table I). The mechanism of ion-conduction is described by the fractional

| Dose (kGy) | 0 | 20 | 60 | 80 |
|---|---|---|---|---|
| Thickness (µm) | 463 | 566 | 556 | 398 |
| $\lambda$ Debye length (µm) | 2.051 | 0.586 | 0.738 | 0.760 |
| Relative dielectric constant ($\varepsilon_r$) | 340 | 24 | 38 | 42 |
| $q$ (order of fractional derivative) | 0.789 | 0.73 | 0.76 | 0.73 |
| $h$ (ratio of normal to anomalous diff. contr.) | 0.678 | 0.805 | 0.740 | 0.740 |

| Dose (kGy) | 0 | 20 | 60 | 80 |
|---|---|---|---|---|
| $\mathbb{D}_N$ (m²s⁻¹) (10⁻⁵) | 2.245 | 0.0305 | 0.158 | 0.0045 |
| $\mathbb{D}_A$ (m²s⁻¹) (10⁻⁷) | 0.50 | 0.0086 | 0.034 | 0.027 |
| $\kappa_N$ (ms⁻¹) (10⁻⁹) | 21.25 | 21.25 | 21.25 | 21.25 |
| $\kappa_A$ (ms⁻¹) (10⁻⁹) | 0.85 | 0.85 | 0.85 | 0.85 |
| $\tau_N$ (s) | 0.8032 | 0.8032 | 0.8032 | 0.8032 |
| $a$ (s) | 0.82 | 0.92 | 0.818 | 0.638 |
| $\overline{q}_0$ (fractional order of relaxation time) | 0.71 | 0.71 | 0.71 | 0.71 |

diffusion equation with order $q$. The value of $q$ may however vary with plasticizer as well as dose rate. We have allowed $q$ to vary in between 0.73 to 0.79 which is considerably less than 1, indicating the importance of anomalous diffusion processes. Anomalous diffusion in found in the sample before as well as after radiation. The results are highly sensitive to q and the fits in Fig. 4 worsen considerably if $q$ is set to a common value for all doses.

It is very clear that the sharp decrease of diffusion coefficients $\mathbb{D}_N$ and $\mathbb{D}_A$ after irradiation is compatible with the decrease of DC conductivity which is manifested in Fig. 2. Now all parameters are adjusted to give a best fit to the experimental curves for Real Z and Imaginary Z as function of frequency. The Debye length is sharply decreased by one order after irradiation and then monotonically increases for further increase in dose rate. The charge carrier concentration N in the expression for Debye length $\lambda$ is the sum of the ion concentration for x=0 and the additional ions for nonzero x for different doses [11, 30]. In the present work we choose,
$N= 0.5 \times 10^{20}$ m⁻³.

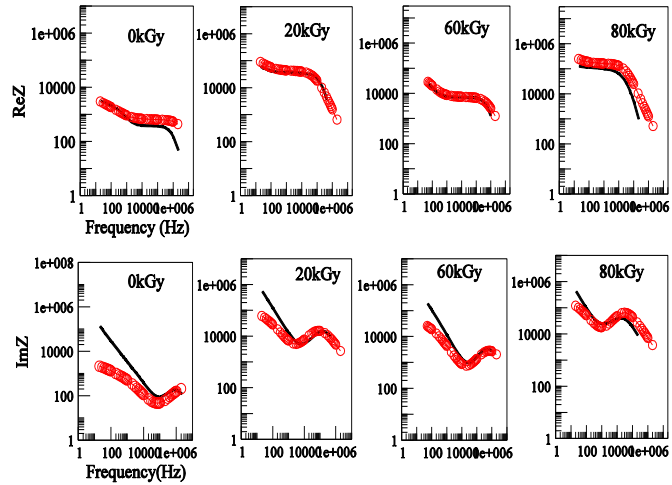

Fig.4 1st row ReZ vs. Frequency. 2nd row ImZ vs. Frequency (solid black line for theoretical value, red circles for experimental value). The experimental and theoretical curves match exactly in some cases and appear indistinguishable.

The decrease in the value of the relative dielectric constant $\varepsilon_r$ also corresponds to the decrease of the number of charge carriers after irradiation. $\kappa_N$ (ms$^{-1}$) and $\kappa_A$ (ms$^{-1}$) are the phenomenological parameters [19, 20] for normal and anomalous relaxation times $\tau_N$ (sec) and $\tau_A$ (sec) for the adsorption and desorption processes near the interface of the sample surface and electrode. The effective value of the products $\kappa_N \tau_N$ and $\kappa_A \tau_A$ which have the dimension of length (m) are ~0.0171μm and 0.167nm respectively. These are termed as characteristic lengths, and indicate two different diffusion regions near the surface. The sample thickness is ~ 300-600μm which is much greater than these layers. So the charge situated near the interface of the SPEs and electrodes can exhibit more than one diffusion mode. The exponent $\bar{q}_0$ ~ 0.71 has a fractional value signifying the dynamical behavior of the mobile charges inside the two different layers near the interface responsible. As the thickness of these layers increases the relaxation time will also decrease. The fitted curves for 0kGy, 20kGy, 40kGy and 80kGy are shown in Fig. 4 for 35.71 wt% glycerol.

## 7. CONCLUSIONS

In the present work both the anomalous as well as the normal diffusive regimes are superimposed to get the electrical impedance analytically. For a realistic result matching the experiments, the combined effect of the fractional time derivatives along with the normal that is integer order is required. We have analyzed the behavior of the gamma irradiated sample over a wide frequency range and have successfully reproduced experimental results using the suggested theoretical model. We consider the anomalous diffusion regimes in the mid portion of the sample as well as at the interfaces of the SPEs with the brass electrode. This is achieved by formulating the boundary condition in such amanner that both anomalous and conventional processes are involved. The interface length or the characteristic length obtained from the fitted curves give a physical support to the dynamical behavior of the mobile charges near the interface.

*Acknowledgment*

T. B. thanks to T. Dutta, physics Department, St. Xavier's College, Kolkata for her suggestions. Authors are also thankful to A. Saha, A. Datta, Chemistry Department, Jadavpur University, 2$^{nd}$ campus and P. Bhattacharyay, Food Technology, Jadavpur University for giving the facility of Gamma Irradiation. T. B. thanks to DST for Inspire Fellowship for financial support.